\documentclass[preprint]{aastex}
\usepackage{emulateapj5, psfig, epsfig}
\slugcomment{Accepted by AJ - BAP02-2002-04-OAB}

\shorttitle{The Luminosity Function of NGC~288}
\shortauthors{Bellazzini et al.}

\begin{document}

\title{Deep HST-WFPC2 photometry of NGC 288. II.\\
    The Main Sequence Luminosity Function
    \thanks{Based on observations made with the NASA/ESA Hubble Space
    Telescope at the Space Telescope Science Institute. STScI is operated by the
    Association of Universities for Research in Astronomy, Inc. under NASA
    contract NAS 5-26555. These observations are associated with proposal \#
    GO-6804.}}

\author{Michele Bellazzini, Flavio Fusi Pecci\altaffilmark{1},
Paolo Montegriffo, Maria Messineo\altaffilmark{2}}
\affil{INAF - Osservatorio Astronomico di Bologna, Via Ranzani 1, 40127, 
Bologna, ITALY}
\email{bellazzini@bo.astro.it, flavio@bo.astro.it,
messineo@strw.leidenuniv.nl}

\author{Lorenzo Monaco}
\affil{Dip. di Astronomia, Universit\`a di Bologna, Via Ranzani 1, 40127,
Bologna, ITALY}
\email{s\_monaco@bo.astro.it}

\and

\author{Robert T. Rood}
\affil{Department of Astronomy, University of Virginia,
        P.O. Box 3818, Charlottesville, VA, 22903-0818 }
\email{rtr@virginia.edu}

\altaffiltext{1}{INAF - Stazione Astronomica di Cagliari, Loc. Poggio
dei Pini, Strada 54, 09012 Capoterra (CA), ITALY}
\altaffiltext{2}{presently at Sterrenwach Leiden, Postbus 9513, 2300 RA
Leiden,
The Netherlands}


\begin{abstract}

The Main Sequence Luminosity Function (LF) of the Galactic globular
cluster NGC~288 has been obtained using deep WFPC2 photometry.  We
have employed a new method to correct for completeness and fully account for
bin-to-bin migration due to blending and/or observational scatter. 
The effect of
the presence of binary systems in the final LF is quantified and is
found to be negligible. There is a strong indication of the mass
segregation of unevolved single stars and clear signs of a depletion
of low mass stars in NGC~288 with respect to other clusters. The
results are in good agreement with the prediction of theoretical
models of the dynamical evolution of NGC~288 that take into account
the extreme orbital properties of this cluster.

\end{abstract}



\section{Introduction}

After a decade of HST observations, the Luminosity Function (LF) of
unevolved Main Sequence (MS) stars has been derived for several
Galactic globular clusters (GGCs).  In many cases the LF extends down
to the faintest stellar objects
\cite[$M\simeq 0.1\, M_{\odot}$; see, e.g.,][]{king,bedin}.
For unevolved stars, the luminosity directly tracks the mass, the most
fundamental parameter of a star.  The determination of the
Initial Mass Function (IMF) from the observed LF of Main Sequence
stars, as well as the study of the of the variation of the LF within a
given cluster and among different clusters is now a well established
field of research which has provided important insights on the
dynamical evolution of GCCs \cite[see, e.g.,][and references
therein]{fer97,king,marc,manu00,pdm}.  In particular, the signature of
mass segregation, due to the energy equipartition established by two
body encounters among clusters stars, has been detected in many
clusters by comparing the LFs obtained at different distances from the
cluster center. Also, the effect of the pruning of the outer extremities of
the clusters by the Galactic tidal field or by the violent interaction
with the Galactic bulge and disc have been shown with the
analysis of the LF of MS stars \cite[see, e.g.,][and references
therein]{glo}.

Here we present the MS LF of the globular cluster NGC~288 down to
$V\sim 24.5$. The LF was based on HST-WFPC2 observations that are
described in detail in a companion paper \cite[][hereafter Pap
I]{miki}, that was devoted to the study of the binary population of
this cluster.  In Pap I we used a method very similar to the one
adopted by \citet{ruba1}, based on extensive sets of artificial star
experiments, to estimate the global binary fraction ($f_b$) of the cluster
($ 7$ \% $\le f_b\le 30 $ \% ) and to show that binary systems are spatially
segregated within the cluster. These same artificial star experiments
can be used to obtain LFs in different regions of the cluster, {\em
corrected for all observational effects}. In \S2 we will briefly
recall the characteristics of the adopted sample and of the artificial
star experiments, as well as the results of Pap I that are relevant in
the present context. In \S3 we introduce a {\em new method} to correct
the observed LF. The effect of the presence of binary systems on the
final LF is also discussed and quantified with the support of the
analysis performed in Pap I.  In \S4 we present the final LF of
NGC~288 and compare it with the LFs of other clusters and, finally, in
\S5 we summarize our results.

\subsection{The globular cluster NGC~288}

NGC~288 is a loose 
\cite[ $ \log \rho_0 = 1.80~ L_{\odot V}\,pc^{-3}$;][]{dj93} cluster. 
It is a classical old GGC \citep{ros99}
with a blue Horizontal Branch (HB) morphology \cite[see][and
references therein]{mike,cat}.  Its stars are not exceedingly metal
deficient (${\rm [Fe/H]}=-1.39$) and have the abundance pattern typical of
halo stars \citep{sk00}.  The cluster lies on a very inclined and
eccentric orbit that carries it into the inner, denser region of the
Galaxy, where the effects of Galactic tides and disc/bulge shocks are
more disruptive .  According to \citet{dana} the perigalactic distance
is $R_p\simeq 1.8$ kpc, the eccentricity is $e\simeq 0.7$, the orbital
period is $P\simeq 220 \times 10^6$ years, and the ``destruction
rate'' is one of the highest in the GGC system. The observed distance from
the Galactic Center \cite[$R_{GC}\simeq 11$ kpc;]{harris} indicates
that the cluster is presently near the apogalactic point of its orbit
\cite[see also][]{anna}.

\section{The dataset and the artificial star experiment}

The observations, the data reduction process and the artificial star
experiments are described in detail in Pap~I. Here we summarize the points
that are relevant in the present context.

\subsection{The sample}

Two HST-WFPC2 fields were observed in the F555W and F814W passbands:
one with the PC camera nearly centered on the center of the cluster
(Int field), and one with the WF4 camera partially overlapping the
WF3-Int field (Ext field; see Fig.~1 of Pap~I). The Int field is
almost completely enclosed in the region within 1 half light radius
\cite[$1\, r_h \simeq 1.6\, r_c$; 
where $r_c = 85^{"}$ is the core radius)][]{dj93} 
while the Ext
field samples a region contained in the annulus $1\, r_h< r < 2\, r_h$,
where $r$ is the distance from the cluster center.  The data were
transformed to the standard Johnson-Kron-Cousins photometric system by
calibration to the ground-based data of \citet{ros00}. The Int sample
contains 5766 sources from $V\sim 13$ to $V\sim 25$ and the Ext one
contains 2013 sources between $V\sim 16$ and $V\sim 25.5$. The large
majority of the observed stars are fainter than the Turn Off (TO)
point ($V\simeq 19$), and hence are unevolved MS stars (and/or binary
systems containing {\em two} unevolved MS stars; see Pap~I). The F336W and
F225W observations of the Int field that has been shortly discussed in Pap~I
are not used in the present paper.

\subsection{The artificial star experiments}

Simulations containing more than 80,000 artificial stars per chip
have been made on the $V,~I$ frames. In each run of the artificial stars 
experiments $\simeq 100$ artificial stars per chip were added at the same 
position in the median image (that we used as a master-frame to obtain a list 
of positions of {\em bona fide stars}; see Pap~I) and in each of the single 
frames (that were used to obtain accurate photometry; see Pap~I). The set
of experiments for a given chip (sub-field) was completed with $\sim 800$
independent runs.
The final total number of
artificial star experiments is $> 1,500,000$. For each artificial star
a random input $V$ magnitude is drawn from a distribution similar to
the observed LF extrapolated to magnitudes slightly below the observed
lower limit. This fainter limit was required so that the simulations
would contain a large number of stars fainter than the observed
limiting magnitude to correctly sample the magnitude bins where the
incompleteness is expected to be highest (see Pap~I).

We ultimately derive the
true LF in a way that does not depend on the assumed artificial star
LF (see \S3.2).  The input $I$ is obtained from the input $V$ using the cluster
ridge line. Each simulated star thus had the appropriate color and was
simultaneously added to the corresponding $V$ and $I$ frames at the
same position.  The simulated stars were randomly distributed on the
frames with an additional constraint \cite[see Pap~I and also][for a
detailed description of the method]{tosi} preventing any interference
between simulated stars. In practice, for each run of the artificial stars
experiment, only one artificial star is added in each $80\times80$ pixel 
portion of a chip.
The whole process of data reduction has been
repeated on all the frames with artificial stars added and the {\em
output} magnitudes have been recorded (see Pap~I, for details).

\begin{figure*}
\figurenum{1}
\centerline{\psfig{figure=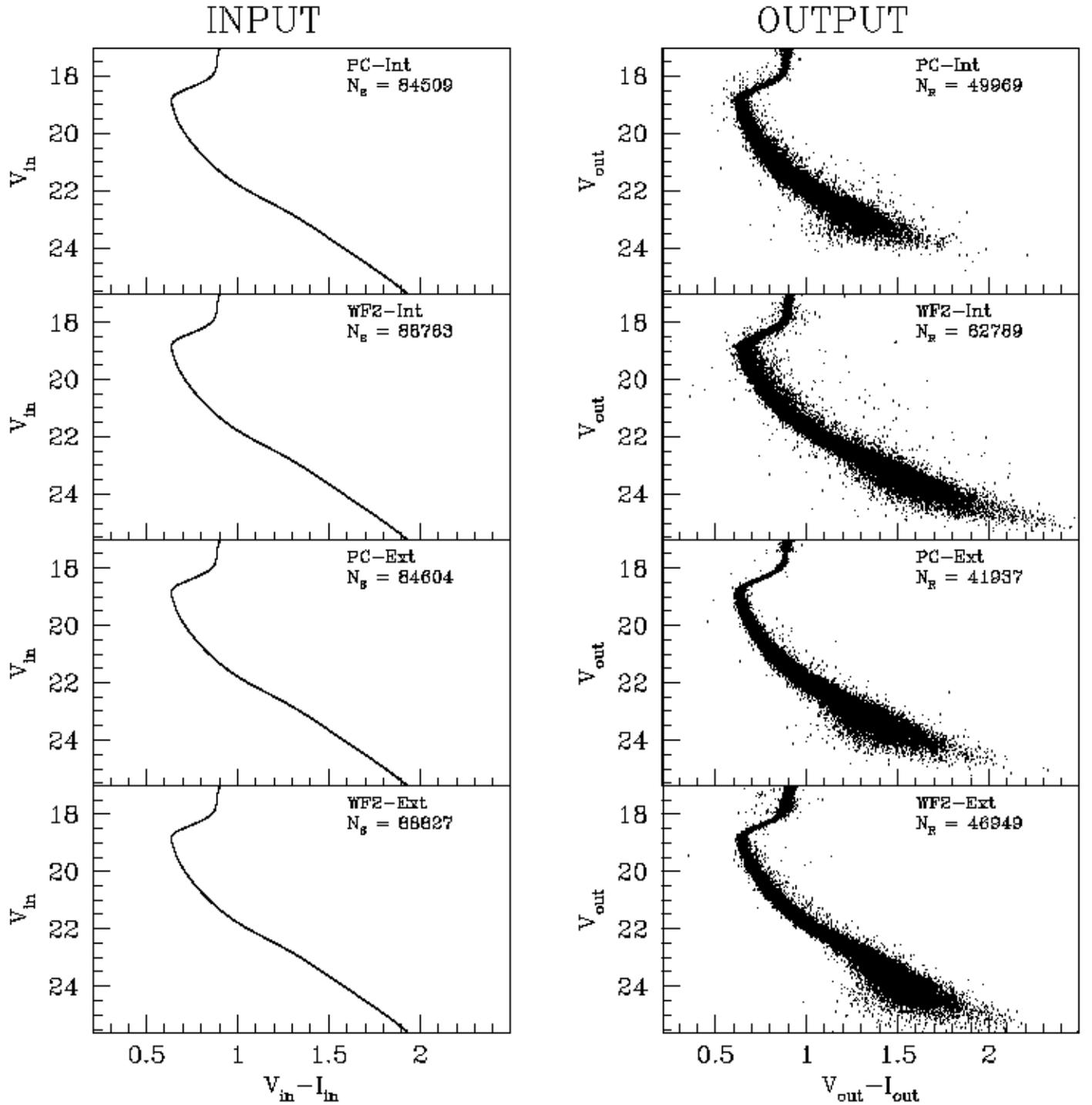}}
\caption{CMDs of the simulated stars obtained with the {\em input}
magnitudes ($V_{\rm in},I_{\rm in}$; left panels), and with the {\em
output} magnitudes ($V_{\rm out},I_{\rm out}$; right panels). The plot
shows the effect of the observation - reduction process on the PC-Int
and PC-Ext frames, while WF2-Int and WF2-Ext are shown as
representative of the WF cameras in the Int and Ext samples,
respectively. $N_S$ is the total number of {\em simulated} stars while
$N_R$ is the total number of {\em recovered} stars. Note that the
ratio between this two numbers is not directly comparable, at face
value, since the limiting magnitude of the LF from which the
artificial stars were drawn was different in the different cases.
Since the observed limiting magnitude was different in the various
cases, the limiting magnitude of the artficial stars LF was varied
accordingly to exclude unnecessarily faint stars (e.g. more than $\sim 1$
mag below than the observed limiting magnitude) from the simulations.}
\end{figure*}

Fig.~1 shows the {\em input} and {\em output} CMDs for the stars
simulated in the PC-Int, PC-Ext, WF2-int and WF2-Ext frames. The WF2
samples have been chosen as representative of the WF cameras for the
Int and Ext fields. Within a given field (Int and Ext) the
observations with the different WF cameras have very similar
properties (see Pap~I).  Note the different response to the same input
color-magnitude distribution in the four cases represented. This
arises because the larger pixels of the WF cameras allow them to reach
fainter limiting magnitudes than the PC camera and because the Int and
Ext fields have different average degrees of crowding.

All of the effects of the whole process of {\em observation and data
reduction} on a large sample of stars representing a pure Simple
Stellar Population (SSP) are shown in Fig.~1. These effects can be imagined
to be the results of the application of a multi-dimensional function
(that we will call {Observation/Measure Function}; hereafter OMF) on
the {\em true} stars of the clusters, here represented by the input
artificial stars lying on the ridge-line. In Pap~I the empirical
knowledge of the effects of the OMF has been used to extract the {\em
signal} provided by the presence of a significant population of binary
systems in the color distribution around the ridge-line from the
observational noise due to random scatter, blending,
etc. Here we will use it to correct the observed LF from all the
spurious deformations due to the same factors.

\subsection{Completeness and division in subsamples}

The artificial star experiments have been carried out independently
for each camera of each observed field. The completeness functions for
each camera have been presented in Fig.~5 of Pap~I. The crowding
conditions of the dataset are never particularly critical, and the
completeness is rather similar everywhere.  The completeness factor
($C_f$) is larger than 80 \% for $V\le 23.5$ in all the samples
except the PC-Int, the innermost one, where $C_f\ge 80$ \% for $V\le
22$. It has been shown that the spatial variation of completeness
within a single camera are negligible. For the present purpose, we
take advantage of the great similarity of crowding conditions in the
WF cameras of the Int and Ext samples to subdivide the total sample in
the following subsamples: (1) PC-Int, (2) WF-Int, containing the
WF2-Int, WF3-Int and WF4-Int samples, (3) PC-Ext and, (4) WF-Ext,
containing the WF2-Ext, WF3-Ext and WF4-Ext samples. The four
subsamples are internally very homogeneous and cover different radial
regions of the cluster. In this sense, the difference between the
PC-Ext and the WF-Ext samples is small but we prefered to keep them
separate because of the different limiting magnitude.

\section{The correction of the observed LF}

The effects of the OMF on the {\em true} properties of stars that may
affect the LF derived from an observed sample are:

\begin{enumerate}

\item {\bf Star loss.} Stars may be lost in the observation/measure process.
This occurs for two different reasons:

\begin{enumerate}

\item Near the limiting magnitude of the observations a star can be
succesfully detected if it falls on a positive fluctuation of the
background that brings it {\em above} the detection threshold and can
be {\em lost} if it falls on a {\em negative} fluctuation. This effect
is purely stochastic; it does not depend on the crowding. It
just depends on the amplitude of the random fluctuations of the
background. There is a level of magnitude beyond which no possible
positive fluctuation can bring a star above the detection
threshold. All the stars fainter than this limit are lost (unobserved)
and just contribute to the diffuse luminosity of the unresolved
background.

\item Stars may be lost if they fall on a brighter source (star or galaxy)
and/or if their images are badly corrupted by defects of the chip or
the hit of a cosmic ray (this two latter effects are expected to have
a negligible impact in the present case because of the adopted
observation + data reduction strategy; see Pap~I). This effect depends
on the local degree of crowding as well as on the magnitude of the
star.

\end{enumerate}

\item {\bf Blending.} A star may be blended with a fainter source and
its measured magnitude will be brighter than the {\em true} one
because of the flux contributed by the fainter blended
source. Considered from the perspective of artificial stars
experiments the maximum possible effect of a blending is to change the
observed magnitude of a star by $\simeq 0.75$ mag. In fact, if we
recover an artificial star by its (known) position in a frame with an
output magnitude more than $0.75$ mag brighter than its input
magnitude, this means that it is fallen on a real star brighter than the
simulated star. Thus, we are not recovering/measuring the simulated star 
but the real one instead, i.e., thus the artificial star has been lost.

\item {\bf Observational scatter.} The magnitude of the {\em true}
star is altered by all the sources of stocastic noise associated with
the process of observation/data reduction (e.g., photon noise,
imperfect PSF fit, etc.).  The maximum possible effect of this factor
is magnitude dependent and may be roughly evaluated on the basis of
the typical error on a given measure. For instance, in the present
case we included in the final samples only the stars with errors\footnote{The
error associated to each magnitude entry is the error on the mean over repeated
measures, see Pap~I.} lower than 0.2 mag in both $I$ and $V$. 
Hence a $\sim 3\sigma$ fluctuation can change the magnitude of an observed star 
by $\sim 0.6$ mag, at most, near the limiting magnitude level.

\end{enumerate}

The derived LFs are in practice histograms of the number of stars as a
function of magnitude. The factors above change the {\em true} LF
histogram by depopulating the bins as a function of (increasing)
magnitude (mainly factor 1) and by moving stars from one bin to
another (migration) due to the magnitude changes induced by factor 2
and/or 3.  However factors 2 and 3 are strictly interwoven and their
effect cannot be separated. For example, a given star may be moved,
say, 0.4 mag brighter because it is blended with a fainter unresolved
source, but it might simultaneously be affected by a negative
fluctuation of the random observational scatter, so its final
magnitude may differ from the {true} one by just 0.2 mag. A blending
between two unresolved sources can bring them above the detection
threshold, injecting in the sample two stars that should have been
lost if not blended\footnote{Note that this phenomenon may
significantly alter the star counts in the bins near the limiting
magnitude, at least in cases in which the crowding conditions are
critical, see \citet{tosi}, for discussion}. The possibilities are too
numerous to list here.  The real contribution of the three effects
cannot be known for a the specific case. On the other hand, it is
clear that we can obtain a statistical knowledge of the actual
LF.

The traditional way to recover the {\em true} LF from an observed one
consists in multiplying the numbers in each bin of the observed LF by the
inverse of the Completeness factor ($C_f=N_R/N_S$; where $N_R$ is the number of
{\em recovered} stars and $N_S$ is the number of the {\em simulated} ones) 
at the center of the bins. 
It is clear that such procedure corrects at first order only
for the effects of star loss (factor 1) while the effects of the other
factors are ignored. A popular misconception is that the observational
scatter cannot produce bin migration if the width of the bin is larger
than the typical error on the estimate of the magnitude.  This is
obviously not true since a star whose true magnitude is near the edge
of a bin can migrate because of an arbitrarily small fluctuation. The effects of
migration may be amplified by the completeness correction, especially in the
faintest bins of the LF.

Furthermore the Completeness factor may be obtained from the artificial star
experiments {\em only as a function of input magnitude} (say, $V_{\rm in}$) while
the observed LF is (obviously) a function of the {\em observed magnitude}, whose
corresponding quantity in the artificial star experiment is
$V_{\rm out}$\footnote{The input magnitude ($V_{\rm in}$) of artificial stars 
corresponds to the {\em true} magnitude of real stars ($V_{\rm true}$), while the
output magnitude ($V_{\rm out}$) corresponds to the observed one ($V_{obs}$) under
the extant conditions}. Thus when an observed LF is corrected by multiplying it
by $1/C_f(V_{\rm in})$, it is implicitly assumed that $V_{\rm in}$ and $V_{obs}$ are the
same quantity, that, in general, is not true.

There have been few attempts to correct the observed LFs for all the
effects above \citep{druk,stha}. The proposed methods require a complex
mathematical approach, and may rely on the {\em a priori} assumption of an
analytical form of the true LF \citep{stha} or on approximate treatments of
the propagation of errors under matrix inversions 
\cite[see][for references and discussion]{druk,ken}.

Here we propose a new method that is fully non-parametric and extremely simple
in nature, taking also advantage of the huge enhancement in computational power
of workstaions and PCs that took place since the times of the above quoted
attempts. As we will show below
({\em a}) the proposed method takes full account of the effects of bin 
migration and ({\em b}) the completeness correction is consistently applied to a
LF that is function of $V_{\rm in}$ as the completeness factor.
The method is particularly appropriate for the application
to LFs of SSPs, but it can be probably generalized to other cases. The
prerequisite is the availability of huge sets of realistic artificial
star experiments well sampling the whole spatial, luminosity, and
color range of the observations.

\subsection{The Equivalent Sample}

Consider a large set of artificial star experiments performed in a
region of uniform crowding conditions, e.g., the WF2-Ext set presented
in Fig.~1. Suppose that the input magnitudes of the artificial stars be
distributed as the ``true'' stars of the parent population (in this case the
stars of NGC~288; the reasons of this hypothesis will become clear at the end 
ofthis section). 
From this set it is possible to extract a subsample of {\em
recovered} stars with exactely the {\em same dimension} and the 
{\em same distribution as a function of $V_{\rm out}$} as the observed sample. 
Call this subsample an {\em Equivalent Sample} (ES)\footnote{The operational 
procedure to extract an ES is the following. Consider the
histogram representing the {\em observed} LF: each bin {\em i} contains
$N_i$ stars. From the artificial star set we extract at random $N_i$ stars 
whose {\em output} magnitudes lie in the {\em i}-th bin. Repeating this step for
each bin of the observed LF we obtain a sample of recovered artificial star 
that have an {\em output} LF equal to the {\em observed} one. Note that the
same procedure can be applied on a star-by-star basis, by randomly extracting 
an {\em artificial sister} (see \S3.4) for each {\em observed} star.}. 
It is important to note
that the ES contains only stars that have been succesfully recovered in the
artificial star experiments process, hence that have suffered
only  the effects of bin migration (i.e, blending and/or observational scatter).

By definition the ES has {\em the same output 
(e.g., observed) properties} of the observed sample, it is one of the many
possible relizations that are mapped by the OMF from the input/true domain to
the output/observed one. Thus {\em the input (e.g., true)} properties of the ES
are one of the possible {\em true} set of properties that may have been mapped 
into the observed sample by the OMF. 
It is immediately apparent that the LF of 
the input magnitudes of the ES is equivalent to the observed LF 
{\em once corrected for all the effects of bin migration}\footnote{In other
words the LF of the input magnitudes of the ES is the observed LF seen 
{\em before} any blending and/or observational error take place.}. Thus the
identification of an ES allows ({\em a}) the removal of all the effects of bin
migration and ({\em b}) the expression of the observed LF as a function of input
magnitude, consistently with the completeness factor (see \S3.).
If we apply the completeness correction to the {\em input LF} of the ES we find 
one possible realization of the final LF, corrected for {\em all the 
observational effects}. 
Since different ES may be mapped into the same observed sample by the
OMF, repeating the process for a number of (randomly extracted) different ES 
and taking the average will allow to obtain a final LF more robust to
random fluctuations as well as a direct estimate of the uncertainty associated
with the whole process.

However, there is another important factor we have to take into
account. While the completeness correction is completely independent
from the distribution in magnitude of the artificial star set, the
derived correction for bin migration is not. Consider a given bin of
the observed LF. Such a bin will be primarily filled by stars with
$V_{\rm out}\simeq V_{\rm in}$, but also by a certain fraction of
stars which have migrated from the nearby bins beacause of blending
and photometric errors.  Increasing the number of artificial stars in
the nearby bins, with respect to the considered bin, will enhance the
probability of extracting stars that enter in the bin because of
migration, hence increasing the final ``migration correction''. The
right correction would be obtained only if the artificial star set is
distributed as the true LF of the population which is unknown, being
the very target of our analysis. We circumvent this problem by
iteratively adjusting ({\em a posteriori}) the distribution of the
artificial star set until the best approximation is obtained (see
below for details).  The approach has proven to be successful since we
obtain the same final LF {\em independently} of the initial
distribution of the artificial star set. In the following section the
adopted operational procedure is described in detail.

\subsection{Operational procedure}

For each of the four observed subsamples
(PC-Int, WF-Int, PC-Ext, WF-Ext) we adopted the following procedure:

\begin{enumerate}

\item An initial distribution of input magnitudes is assumed for the artificial
      stars and the largest possible subset accordingly distributed is extracted
      from the whole set of artificial stars.

\item Twenty different ES are extracted at random from the above subset and
      their $LF(V_{\rm in})$ are corrected for completeness. 

\item The (bin per bin) average of the twenty obtained LFs is adopted as the
      ``current corrected LF'' and the standard deviation is adopted as the
      corresponding uncertainty.

\item The ``current corrected LF'' is assumed as the distribution of the
      artificial stars and a new subset of artificial stars
      accordingly distributed is extracted from the whole
      set of artificial stars as in step 1. 

\item The steps 2,3 and 4 are repeated twenty times. The twentieth ``current 
      corrected LF'' is adopted as the {\em final LF} with the corresponding
      uncertainties.

\end{enumerate}

Step 4 ensures that at each new iteration the adopted set 
of artificial stars will become more and more similar to the ``true''
one, giving increasingly appropriate rates of bin migration.

In all the cases considered here the process converged in $\le 6$
iterations, e.g., the ``current corrected LFs'' become very stable
after the sixth iteration. Furthermore, the result is independent of
the initial distribution of input magnitudes that is
assumed for the artificial star subset.  We have tried many different
distributions (some of which are shown in Fig.~2) and in all cases the
process converged to {\em the same final LF} in few iterations.

Some examples of the procedure are shown in Fig.~2, applyied to the WF-Ext
sample. In Fig.~2a the artificial star subset was initially distributed as the
observed LF down to $V\le 25.5$ and uniformly from this limit down to $V=27$.
A scaled version of the adopted distribution is plotted as a thick dotted line.
The thin dotted histogram shows the ``current corrected LF'' after the first
iteration of the process, the thin continuous histogram shows the ``current 
corrected LF'' after the sixth iteration and the open dots are the final LF,
corresponding to the twentieth iteration of the process, with the currrent 
standard deviation of the twenty realizations of the corrected LFs as errorbars.
The encircled dot is the first bin of the LF for which $C_f < 0.5$, i.e. the
completeness correction is larger than a factor 2.
The fast convergence and the stability of the final solution can be readily
appreciated. 

Fig.~2b and 2c have the same symbols of Fig.~2a but shows the convergence of the
process starting from (very) different initial distributions of the artificial 
star subset. The open dots are {\em the same final LF shown in Fig. 2a}. Thus
Fig.~2a and 2b shows that the process converges to {\em the same final LF}
independently of the assumed initial distributions of the artificial stars
subset.

Finally, Fig.~2d shows in detail the results of the twentieth (i.e. final)
iteration of the process shown in Fig.~2a. The twenty histograms are the
corrected LFs obtained by the twenty different ES extracted from the finally
adopted subset of artificial stars. The full dots and errorbars are their means
and standard deviations.

\begin{figure*}
\figurenum{2}
\centerline{\psfig{figure=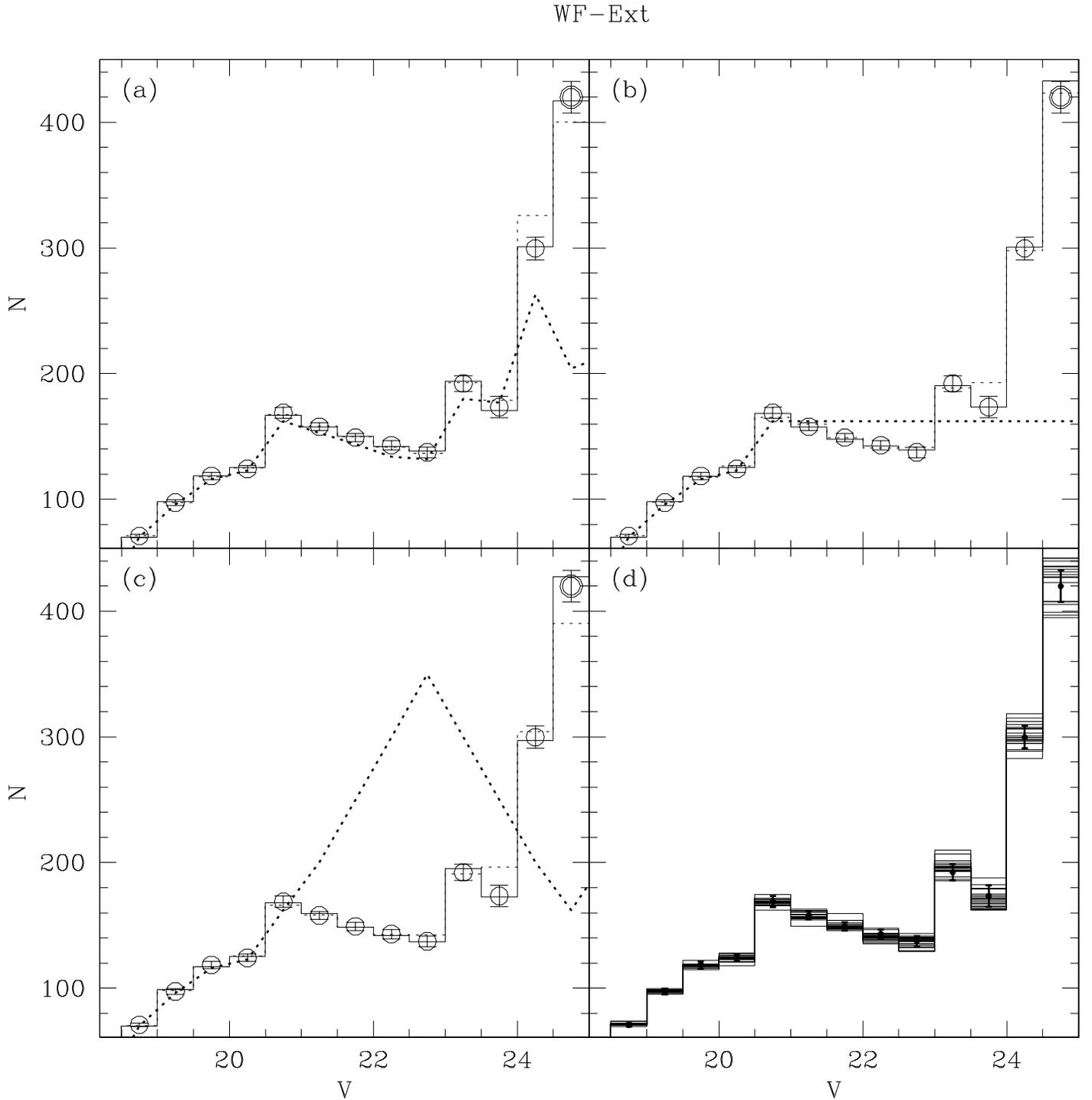}}
\caption{Panel (a): The final LF is shown as open dots with error
bars (the same LF is reported with the same symbols also in panels {\em b} and 
{\em c}). The encircled dot is the first point with  $C_f < 50$ \%. 
The thick dotted line is a scaled version of the distribution of $V_{\rm in}$ 
of the artificial stars subsample adopted for the first iteration, in the 
present case assumed with the same shape of the {\em observed} LF to $V=25$ 
and with $N=215$ for the $V>25$ bins (in the present scale). 
The thin dotted histogram is the LF 
recovered after the {\em first} iteration of the process. The thin continuous 
histogram is the LF recovered after the {\em sixth} iteration, when convergence
has been reached.   
Panel (b): the same as Panel (a) but
adopting a different distribution for the artificial stars subsample adopted for 
the first iteration (i.e., flat for $V\ge 20.5$). Panel (c): the same as panels
(a) and (b) for an initial distribution for the artificial stars subsample
strongly peaked at $V=22.75$. Panel (d): the twenty realizations of the
last iteration of the process for the case shown in Panel (a). The final LF is
the mean of the twenty realizations (full dots) and the uncertainty associated
with the process is the corresponding standard deviation. }
\end{figure*}

\subsubsection{The final LFs}

Since the obtained LFs are independent from the initial distribution
of the adopted subset of artificial stars, we derived our final
(corrected) LFs for all the considered samples assuming the same
initial distribution of the artificial stars shown in Fig.~2a. The
final LFs are given in Table~1 (PC-Int and WF-Int samples) and in
Table~2 (PC-Ext and WF-Ext samples).  The table gives the $V_{\rm in}$
(e.g., $V_{\rm true}$) of the center of the bins and, for each sample,
the observed LF, the completeness factor $C_f$, the final LF and the
total error on the final LF (see below).

The final LF is given down to the first bin that has $C_f<0.50$. In
the following plots this point is circled to serve as a reminder that
it is has been derived applying a large correction. We retain these
points as educated guesses of the behaviour of the LFs beyond the
range in which moderate and safe corrections can be made.  We consider
as fully reliable only the points of the LF for which $C_f\ge0.50$.
We note that the last fully reliable points, according to the above
criterium, have $C_f=0.607,0.769,0.806,0.842$ for the PC-Int, WF-Int,
PC-Ext, and WF-Ext samples, respectively. Thus the completeness is
quite high for the whole considered range.

The final total error reported in Tab.~1 and 2 has been obtained by 
summing in quadrature the error on the Completeness factor 
\citep[estimated according to Eq.~18  by][]{ken}, to the standard deviation of 
the twenty different realizations of the LF that are obtained in the last step 
of the procedure (see \S3.2 and Fig.~2d).
In all the following plots the errorbars associated with the LFs of NGC~288 
represent this final total error.

The final LFs (open dots) are shown with the corresponding observed LF (full
dots) in Fig.~3.

\begin{figure*}
\figurenum{3}
\centerline{\psfig{figure=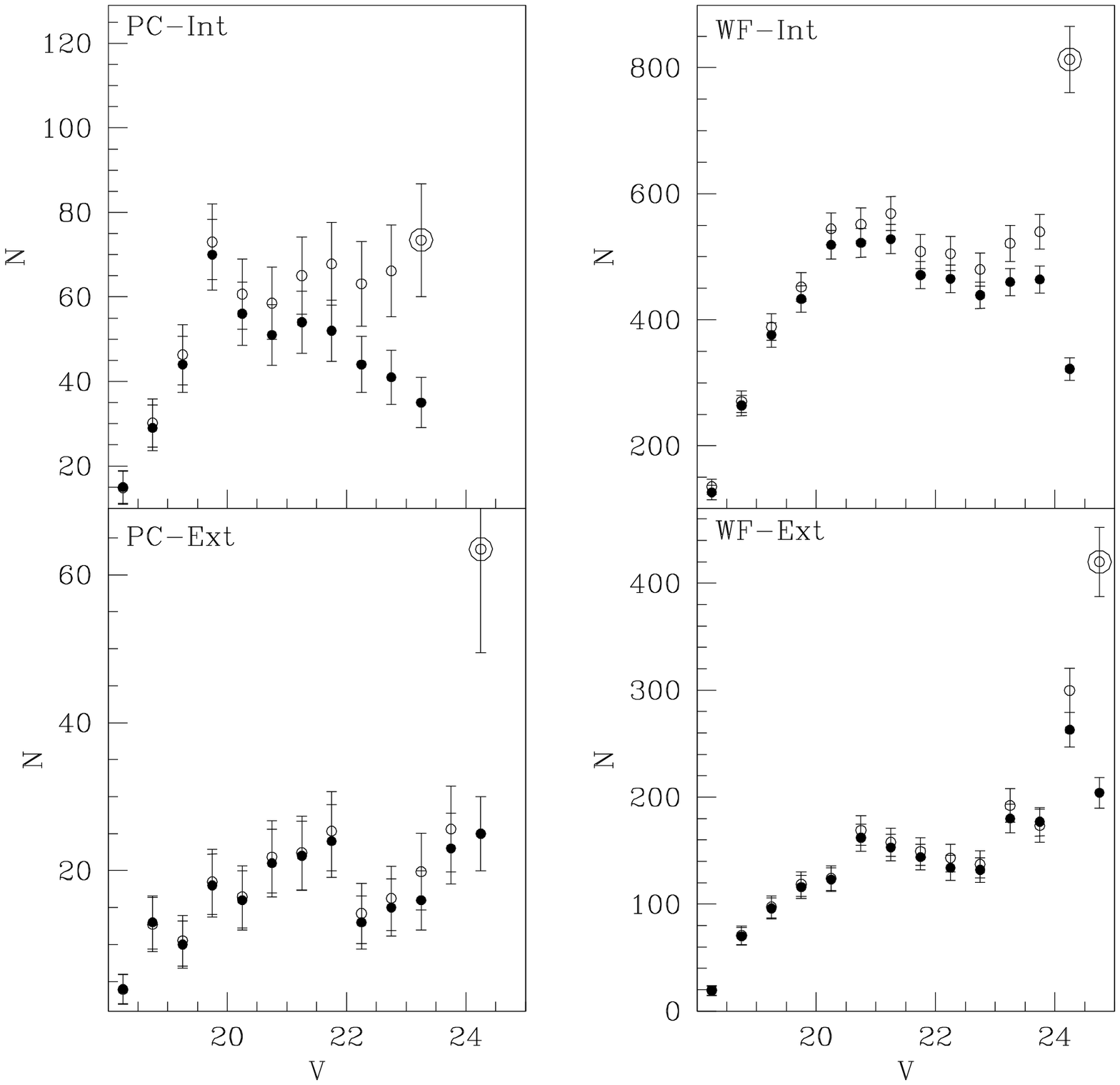}}
\caption{The observed (filled circles) and corrected LFs of the four
samples. The encircled point is the first bin for which $C_f<0.50$.}
\end{figure*}

As a further check we compared our final LFs with those obtained by dividing 
the observed $LF(V_{obs})$  by $C_f(V_{\rm in})$.
We note that the differences are within $1
\sigma$ in most bins and $< 2 \sigma$ in any bin. As expected, in this
particular application the effects of bin migration are not large. 
This would not be the case if much more crowded fields were considered
\citep[see \S3. and][]{tosi}.

\subsection{The effect of binary systems on the LF}

There is an intrinsic kind of blending that cannot be directly tackled
with artificial star experiments. Physical binary systems are strictly
equivalent to chance superposition blendings. The LF for {\em single
stars} cannot be obtained if an independent estimate of the incidence
of binary systems is not available. Further, since the actual binary
fraction ($f_b$) depends on the (unknown) distribution of mass ratios
($F(q)$; see Pap~I), any correction would be model dependent. Indeed,
a direct estimate of the impact of binary systems on the LF of a
globular cluster has only been possible for NGC~6752 \citep{ruba2}.

Now we can add a second cluster using the results of Pap~I. We
measured $f_b$ in the Int and Ext regions of NGC~288.  In the Int region we
found that the observations are compatible with 8\%$\le f_b \le$ 38\%,
and the most probable range is 10\% $\le f_b \le $ 20\%. In the
Ext region $f_b\le$ 10 \% and the most probable value of $f_b$ is
zero, independently of the assumed $F(q)$. Thus, a sizeable fraction
of binary systems is present in NGC~288, at least in the Int sample,
and may affect the derived LF of single stars.

To quantify the effect we computed the mean bin-by-bin ratio between the
LFs obtained by  5 ES with $f_b=20$ \% and by 5 ES with $f_b=0$ \%. This
number gives (approximately) the fractional variation of the star content of 
the bins due to binary systems.  Furthermore, dividing the final LF by the
quoted ratio, the LF is corrected for the binary content and mapped
to the ``pure single stars'' case.  In Fig.~4 the final LF of the
WF-Int sample not corrected for the binary content (filled dots with
error bars) is compared with its binary-corrected versions in the
assumption $f_b=20$ \% and for three different $F(q)$ (see caption and Pap~I for
details). The comparison is performed in the magnitude range in which the 
binary estimate of Pap~I has been performed ($20\le V \le 23.5$).

\begin{figure*}
\figurenum{4}
\centerline{\psfig{figure=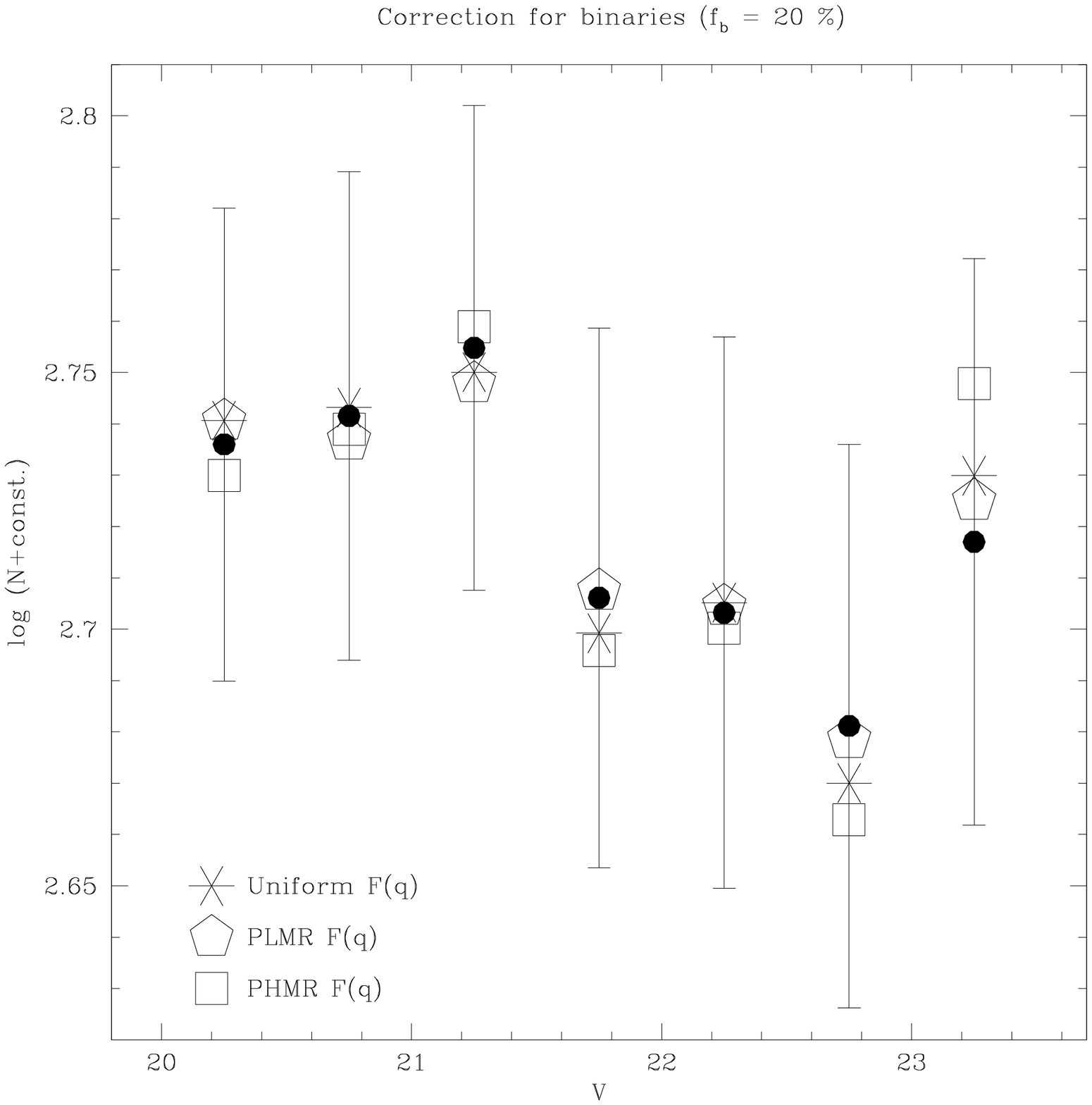}}
\caption{The final LF of the WF-Int sample without correction for
binary systems (full dots with errorbars) is compared with the same LF
after a correction for a binary fraction $f_b=20 \%$, under different
assumption on the mass ratios ditribution ($F(q)$). Open squares: PHMR
(peaked at high mass ratios) $F(q)$; stars: Uniform (all mass
ratios are equally probable) $F(q)$; open pentagons: PLMR (peaked at low mass
ratios) $F(q)$. The corrected LFs are very similar to the uncorrected
one, thus the correction for binary system can be neglected in the
present case.}
\end{figure*}

There are several noteworthy features in Fig.~4: 1) The maximum
correction is $\le $ 3\%.  2) All the corrected LFs lie within the
uncertainties of the un-corrected one.  3) The PLMR $F(q)$ case is most 
similar to the uncorrected case. This was expected, since in this case the
large majority of binary system has a very low mass secondary.  This
means that the migration of the primary is small, and that most of the
secondary stars are so faint that they fall in an unobserved region of
the LF.  4) The correction for the PHMR case depletes the intermediate
bins ($21.5\le V \le 23$; because it corrects for the migration of
primary) and enhances the last bin because of the recovery of some
secondary stars as single stars. 5) No bin is exclusively depleted or
enhanced by the redistribution of the binary components as single
stars.

From the above results and discussion we conclude that:

\begin{enumerate}

\item The correction for binary systems can be safely neglected for
the Ext samples where the binary fraction is much lower than $f_b=20$
\% and possibly null.

\item The correction can be neglected also for the Int samples, at
least for $V< 23.5$, since it is smaller than the uncertainties of the
final LF.  However, it cannot be excluded that the faintest bins of
the single stars LF need a significant enhancement to take into
account the effect of binaries, particularly if the actual $F(q)$ favors
high mass ratios.

\end{enumerate}

\subsection{Further developements and other applications}

The case under study is particularly simple. NGC~288 is a very loose
and relatively nearby cluster, so our WFPC2 observations well resolve
the large majority of the stars down to the limiting magnitude of each
subsample with a very high degree of completeness. Further, the density
gradient is a very fortuitous match to WFPC2 allowing a few large
subsamples to nicely probe radial variations. 

It is important to note that the Equivalent Sample concept does not
require such simplifications and may be successfully applied under any
condition. If a strictly rigorous approach is adopted the ES approach
may have far reaching applications in the study of stellar
populations. Here we provide two possible examples to show the
potential of the method.

\begin{itemize}

\item {\bf Extremely crowded field with strong density gradient}.  The
method can tackle with samples with arbitrarily large density
gradient, for instance by subdividing the observed sample in
subsamples whith nearly uniform stellar density (as done in the
present application).  If the dimension of the observed sample is not
sufficiently large to allow this approach, the ES method can be
implemented also on a star-by-star basis. It will suffice to assemble
the ESs by associating an artificial star analogue for {\em each}
observed star, randomly extracted from a set of artificial stars
having (approximately) the same $V_{\rm out}$ and position in the
frame of the considered observed star ({\em artificial sisters}).  The
only basic requirement to successfully follow this approach is to have
a very large set of artificial stars, sampling with large multiplicity
the whole range of magnitudes and positions covered by the observed
sample.

\item {\bf Mapping synthetic CMDs into the observational plane}. The
standard tool used to derive the Star Formation History (SFH) from the
CMD of a resolved galaxy consist in the reproduction of observed CMD
by a synthethic CMD in which the stars are extracted from theoretical
evolutionary tracks \cite[see][for references and discussion of the
various adopted techniques]{lej}.  To transform the synthetic CMD to
the observational plane the effects of incompleteness, blending and
observational scatter must be added to the synthetic stars. This task
is not easy and it is usually done with simplified approaches. For
instance, the observational scatter is usually introduced by randomly
extracting `errors' from gaussian distributions with standard
deviations equal to the typical $1\sigma$ photometric error. However,
the distribution of the observational scatter is not gaussian, in
general, and it is not easy to parametrize. Furthermore, the effects of
blending are not separable from the effects of observational scatter, and
gaussian distributions do not provide a good reproduction of the
real distribution, introducing undesired noise in the comparisons.

With the ES approach, one can easily map the {\em input magnitudes} (e.g.,
``true'' magnitudes) extracted by the evolutionary track in to the observational
plane with the correct OMF, by ``asking'' to the artificial stars experiments
what is the probabilty that such star would be successfully recovered or not.
If a star passes the ``incompleteness barrier'', it can be correctely mapped into
the observational plane, for instance by assigning to it the output properties 
of an artificial analog extracted at random from a set of 
{\em artificial sisters}.

In this way, the whole effects of the OMF would be correctly reproduced without
dangerous approximations and/or parametrizations.

\end{itemize}

\section{Comparison of LFs}

Once the observed LFs have been corrected for all the
observational effects and the impact of binary systems have been
quantified, we can use the final LF to study the dynamical status
of NGC~288 either comparing the LFs in the Int and Ext regions of the
cluster, either comparing the LF of NGC~288 with those of other
clusters. In the following we will normalize all the samples to be
compared to the number of stars in the range $18.5 \le V \le 20.5$,
following the approach of \citet[][hereafter PZ99]{manu99}.

NGC~288 is not sufficiently nearby to allow a safe derivation of the
LF down to the hydrogen burning limit even with HST. The comparison
with other ``state of the art'' LFs of more nearby clusters (see
Fig.~6, below) shows that such limit would occur a couple of magnitudes
below our faintest valid point (the faintest bin of the WF-Ext
LF). The Mass-Luminosity relation adopted in Pap~I indicates that the
lowest mass efficiently sampled by our deepest LF is $M\simeq 0.35\,
M_{\odot}$.  Usually MF are compared according to their slope in the
plane $\log N - \log M$ for $M\le 0.5\, M_{\odot}$ (see PZ99 and
references therein). Our best LF has just three points below this
limit, thus the slope of the MF would be poorly constrained. Because
of this limitation we avoid any conversion of our LFs to Mass
Functions (MF), since that would necessarily be somewhat model dependent
and uncertain \cite[see, e.g.,][]{bedin}.
Instead we will make direct comparisons between LFs in the allowed range.

\subsection{Mass segregation within single stars in NGC~288}

The LFs of the PC fields are identical (to within the errors) to those of the 
respective WF samples, in the range where the comparison is possible. 
Since the WF samples are much larger and reach fainter magnitudes we will 
limit our analysis to these homogeneous datasets in the following.

\begin{figure*}
\figurenum{5}
\centerline{\psfig{figure=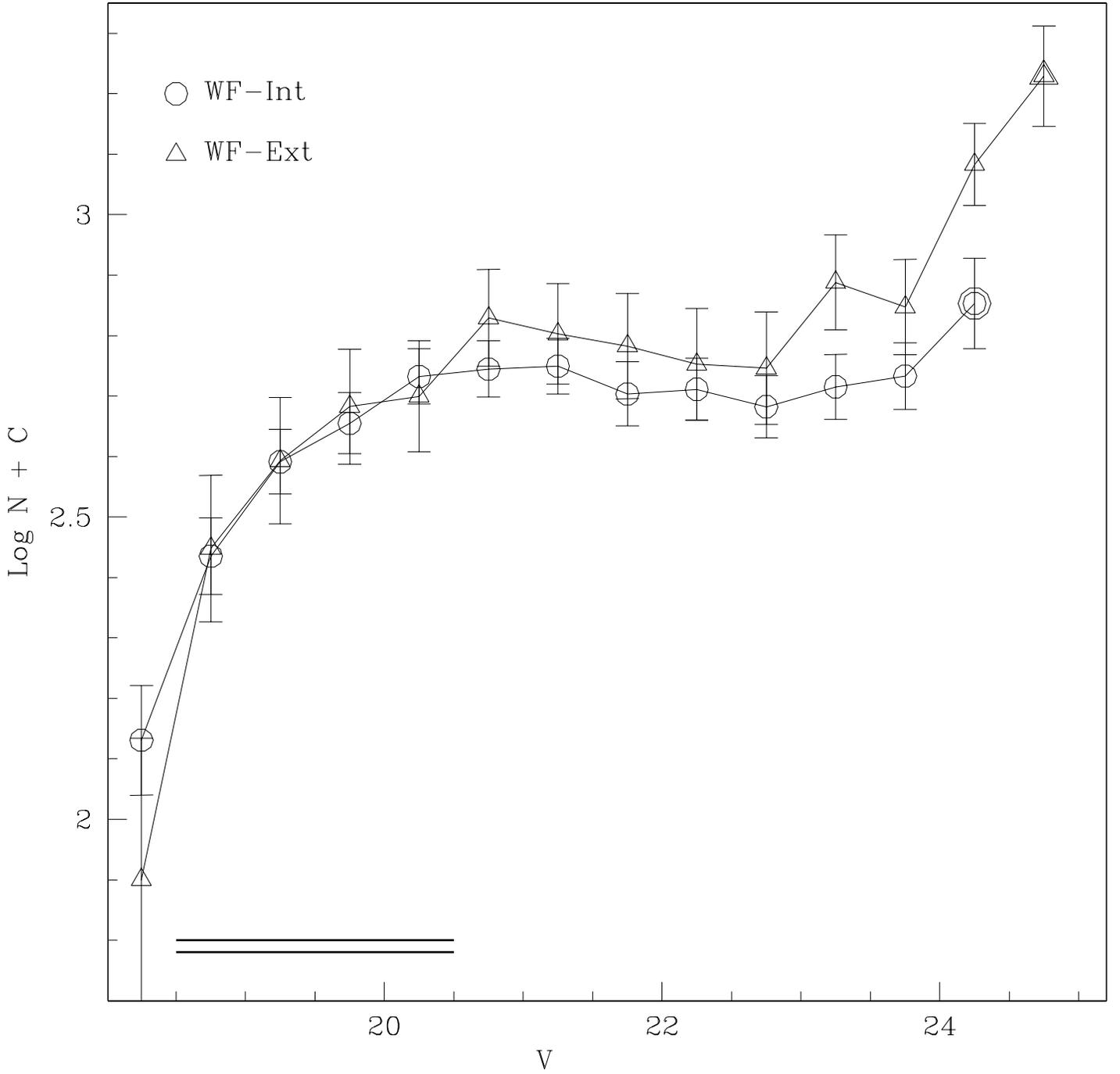}}
\caption{Comparison between the LFs of the WF-Int sample (open circles) and of
the WF-Ext sample (open triangles). The encircled points are the first points 
with $C_f<0.50$.The horizontal bars indicate the normalization interval.}
\end{figure*}

In Fig.~5 the LFs from the WF-Int and the WF-Ext samples are compared.
It is evident at a first glance that the LF of the WF-Int sample is
flatter than that of WF-Ext. Given the adopted normalization, this is
a clear indication of a depletion of low mass stars in the region
within $1\, r_h$ (Int) with respect to the Ext field.   
If the brightest bin is included in the
normalization the effect is significantly enhanced. We note also that any 
correction for binaries would be much smaller than the observed difference (see
Fig.~4).
On the other hand, if fainter stars ($20.5 < V \le 24$) are used for 
normalization, one would infer a substantial excess of the more massive stars 
in the Int sample.

Independently of the assumed normalization it is worth establishing if the 
difference between the two LFs is statistically significant. To check this
hypothesis we performed a $\chi^2$ test, adopting the appropriate definition of
the statistic according to Eq. 14.3.3 by \citet{numrec}. The comparison has been
performed in the safely comparable range $18.5\le V \le 24$. The resulting
reduced $\chi^2$ is 2.31 with 12 degrees of freedom. The difference between 
the two LFs is found to be significant at the $99.1$ \% confidence level, i.e.
{\em highly significant}.   

We have already demonstrated that mass segregation exists in NGC~288,
since both binary systems and their by-product Blue Stragglers Stars
are also more centrally concentrated than the the single
stars. Although the evidence is not as strong (because of the possible ambiguity
in the interpretation associated with the choice of the normalization range), 
we feel that we now find some evidence for further mass segregation in the 
difference between the LFs of unevolved stars in the Int and Ext regions of 
the cluster.

\subsection{Low mass star depletion in NGC~288}

In Fig.~6 the WF-Int and the WF-Ext LFs of NGC~288 are compared with
the LFs of M10, M22 and M55 by PZ99. The transformation from apparent
to absolute magnitude has been obtained by assuming $(m-M)_0=14.67$
and $E(B-V)=0.03$, according to \citet{fer99}. The faintest point of
the WF-Ext LF reaches $M_V\simeq10$ while the LF of the (more nearby) PZ99
clusters reaches $M_V\simeq12$--13. The faintest reported point 
appear to coincide with the turn-over point found in the LFs of the PZ99 clusters, so
this possible feature in the LF of NGC~288 is not detectable in our
data \cite[but it have been apparently detected by][hereafter
PBD00]{anna}.

\begin{figure*}
\figurenum{6}
\centerline{\psfig{figure=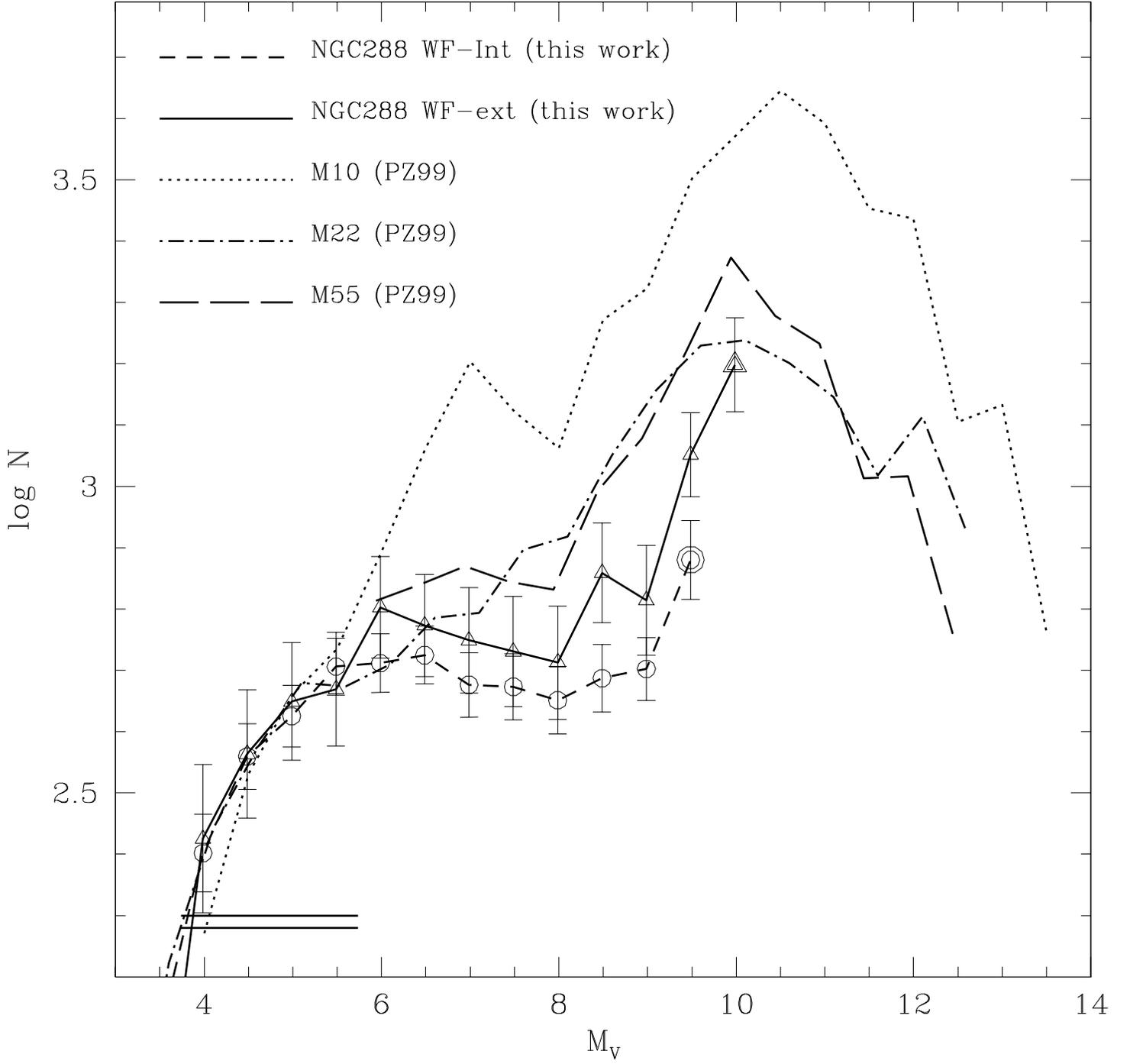}}
\caption{Comparison between the WF-Int and Wf-Ext LFs of NGC~288 with the LFs
of M10, M22 and M55 by PZ99. The encircled points are the first points with
$C_f<0.50$. The horizontal bars indicate the normalization interval.}
\end{figure*}

On the other hand, the comparison in the range $6 < M_V \le 10$ shows
that the LF of NGC~288 is significantly flatter than those of M10,
M22 and M55.  Note that the LFs of these clusters were obtained near
the half-mass radius \cite[usually approximated with the half light
radius $r_h$, see][]{dj93}. This is the region where the mass distribution
of the cluster population is expected to be fairly representative of
the (present day) global population \cite[see][and references
therein]{mh97,vh}. Since the Int and Ext sample are separated at 1
$r_h$, the LF at the half-mass radius must be intermediate between
the WF-Int and the WF-Ext, and the derived LFs should be
representative of the global LF of NGC~288.

The comparison presented in Fig.~6 strongly suggests that the global
population of NGC~288 is significantly depleted of low mass stars with
respect to M10, M22 and M55 (see PZ99 for an indirect comparison
with other clusters).

NGC~288 has a mass ranging from $\sim {1\over{2}}$ to $\sim
{1\over{6}}$ that of the comparison clusters. Its central density
ranges from $\sim {1\over{2}}$ to $\sim {1\over{60}}$ \citep{tad} that
of the comparison clusters.  Thus NGC~288 is intrinsically less
resistent to harrassment and heating by external forces. Furthemore,
among the clusters considered here, it moves on the orbit having the smallest
perigalactic distance, the highest eccentricity and the highest
inclination to the Galactic plane. Hence, it is expected to suffer
strong bulge and disk shocks, and indeed may be one
of the clusters with the highest disruption rates
\citep{dana,oleg}. The coupled effects of internal energy
equipartition (mass segregation), pushing the lighter stars to outer,
shallowly bound orbits, and of the tidal field of the Galaxy,
removing the less bound stars venturing in the outer parts of the
cluster halo, may well be at the origin of the
depletion of lighter stars in NGC~288. We regard this result as a
reassuring confirmation of the theoretical predictions that account
for the orbital characteristics of actual GGCs
\citep{dana,oleg}.

Recently, PBD00 presented a LF in the $J$ and $H$ near-infrared bands
obtained with the HST NIC3-NICMOS camera (parallel observations). The
sample is small but reachs fainter magnitudes (lighter masses) than
our data.  PBD00 conclude that the mass function of NGC~288 is very
similar to that of all other studied clusters, and that it shows no
sign of the strong harrassment predicted by \citet{dana} and
\citet{oleg}.  We regard the results (and the conclusions) of the
present paper as more robust than those by PBD00 since: {\em (a)} the
LF by PBD00 is based on a sample of 75 stars, a very small sample that
may be subject to strong statistical fluctuations, and is not
supported by artificial star experiments, while our LF is based on a
sample of more than $6,000$ stars corrected for all the observational
biases with the artificial star experiments; and {\em (b)} the field
observed by PBD00 is a small stamp ($51\farcs 2 \times 51\farcs 2$,
$\sim 15$ times smaller than the field observed here) located at
$r\sim 2.4\,r_h$, far outside the region of our sample which is
expected to be representative of the global cluster LF.

\section{Summary and Conclusions}

The LF of the globular cluster NGC~288 (down to $M_V\sim 10$) has been
obtained from a field covering a region comprised between the center
of the cluster and $r\sim 2\, r_h$.

A new method to correct the observed LFs for all the
observational effects have been introduced and applied. The method is
based on the Equivalent Sample concept, which may have many
interesting applications in the study of stellar populations.

The effect of the presence of binary systems in the final LF has been
quantified and it has been found negligible, in the considered
magnitude range.

Comparison of the LFs obtained in different regions of the clusters,
indicates the presence of mass segregation within NGC~288. Independently of the
assumed normalization the LFs of the inner (WF-Int) and outer (WF-Ext) sample
turns out to be different at the $99.1$ \% confidence level.

The  comparison
with the LFs of other clusters strongly suggests that the global
population of unevolved stars in NGC~288 is significantly depleted of
low mass stars ($M\le 0.5$--$0.6\, M_{\odot}$), in general agreement
with the predictions of recent theoretical predictions taking into
account also the orbital properties of the cluster.

\acknowledgments

The financial support of the {\em Agenzia Spaziale Italiana} (ASI)
 is kindly acknowledged.
The financial support to M. Messineo has been provided by the Osservatorio
Astronomico di Bologna.  Part of this work has been the subject of the
Thesis of Degree of L. Monaco (Dept. of Astronomy, Bologna
University).  Part of the data analysis has been performed using
software developed by P. Montegriffo at the Osservatorio Astronomico
di Bologna. RTR is partially supported by NASA LTSA Grant NAG 5-6403
and STScI grant GO-8709.  This research has made use of NASA's
Astrophysics Data System Abstract Service.


\clearpage
\begin{deluxetable}{lcccccccccc}
\tablecolumns{11}
\tablewidth{0pc}
\tablecaption{PC-Int and WF-Int Luminosity Functions}
\tablehead{
\multicolumn{1}{l}{} &
\multicolumn{1}{c}{} &
\multicolumn{4}{c}{PC-Int} &   
\multicolumn{1}{c}{} &
\multicolumn{4}{c}{WF-Int} \\
\colhead{$V$} & \colhead{} &\colhead{$N_{\rm obs}$}  & \colhead{$C_f$} & \colhead{$N_c$}
& \colhead{$\sigma_{Nc}$} & \colhead{} &
\colhead{$N_{\rm obs}$} & \colhead{$C_f$}   & \colhead{$N_c$}
& \colhead{$\sigma_{Nc}$}\\}
\startdata
18.25  && 15 & 0.978 & 14.8 & 3.9 && 126 & 0.956 & 135.1 & 12.2\\
18.75  && 29 & 0.963 & 30.2 & 5.7 && 264 & 0.981 & 270.2 & 17.1\\
19.25  && 44 & 0.957 & 46.3 & 7.1 && 376 & 0.975 & 388.6 & 20.8\\
19.75  && 70 & 0.946 & 73.0 & 8.9 && 433 & 0.972 & 452.1 & 22.8\\
20.25  && 56 & 0.920 & 60.7 & 8.3 && 519 & 0.969 & 544.5 & 25.1\\
20.75  && 51 & 0.898 & 58.5 & 8.5 && 522 & 0.956 & 551.5 & 26.2\\
21.25  && 54 & 0.832 & 65.1 & 9.1 && 528 & 0.947 & 568.6 & 26.9\\
21.75  && 52 & 0.767 & 67.8 & 9.8 && 471 & 0.935 & 508.3 & 26.7\\
22.25  && 44 & 0.695 & 63.1 &10.0 && 465 & 0.927 & 504.9 & 27.1\\
22.75  && 41 & 0.607 & 66.2 &10.9 && 439 & 0.913 & 479.9 & 26.3\\
23.25  && 35 & 0.490 & 73.4 &13.3 && 460 & 0.899 & 521.2 & 28.8\\
23.75  && 31 & 0.203 &\nodata&\nodata&& 464 & 0.796 & 539.6 & 27.5\\
24.25  &&  2 & 0.016 &\nodata&\nodata&& 322 & 0.410 & 813.1 & 52.3\\
24.75  &&  0 & 0.000 &\nodata&\nodata&&  75 & 0.096 &\nodata&\nodata\\
25.25  &&  0 & 0.000 &\nodata&\nodata&&   8 & 0.011 &\nodata&\nodata\\
\enddata
\tablecomments{The $V$ columns reports the $V_{\rm in}$ ($V_{\rm true}$) magnitude of 
the center of the 0.5 mag bins. $N_{\rm obs}$ is the observed number of stars 
in the bin. $C_f$ is the completeness factor in the center of the bin.
$N_c$ is the corrected number of stars in the bin (i.e. this column contains the
final corrected LF) and $\sigma_{Nc}$ is the  error on $N_c$. 
The final corrected LFs are reported only down to the first bin with 
$C_f<50 \%$.}
\end{deluxetable}


\clearpage

\begin{deluxetable}{lcccccccccc}
\tablecolumns{11}
\tablewidth{0pc}
\tablecaption{PC-Ext and WF-Ext Luminosity Functions}
\tablehead{
\multicolumn{1}{l}{} &
\multicolumn{1}{c}{} &
\multicolumn{4}{c}{PC-Ext} &   
\multicolumn{1}{c}{} &
\multicolumn{4}{c}{WF-Ext} \\
\colhead{$V$} & \colhead{} &\colhead{$N_{\rm obs}$}  & \colhead{$C_f$} & \colhead{$N_c$}
& \colhead{$\sigma_{Nc}$} & \colhead{} &
\colhead{$N_{\rm obs}$} & \colhead{$C_f$}   & \colhead{$N_c$}
& \colhead{$\sigma_{Nc}$}\\}
\startdata
18.25  &&  4 &  0.996 &  3.9 & 2.0 &&  19 & 0.945 &  19.4 &  4.5 \\
18.75  && 13 &  0.989 & 12.7 & 3.6 &&  70 & 0.991 &  70.8 &  8.6 \\
19.25  && 10 &  0.987 & 10.5 & 3.4 &&  96 & 0.984 &  97.4 & 10.2 \\
19.75  && 18 &  0.980 & 18.5 & 4.4 && 116 & 0.982 & 118.6 & 11.4 \\
20.25  && 16 &  0.980 & 16.4 & 4.2 && 123 & 0.983 & 124.2 & 11.6 \\
20.75  && 21 &  0.980 & 21.8 & 4.9 && 162 & 0.975 & 168.9 & 14.0 \\
21.25  && 22 &  0.964 & 22.4 & 5.0 && 153 & 0.971 & 157.8 & 13.1 \\
21.75  && 24 &  0.947 & 25.3 & 5.3 && 144 & 0.961 & 149.2 & 12.8 \\
22.25  && 13 &  0.915 & 14.2 & 4.1 && 134 & 0.954 & 143.1 & 12.9 \\
22.75  && 15 &  0.895 & 16.3 & 4.3 && 132 & 0.953 & 137.3 & 12.7 \\
23.25  && 16 &  0.879 & 19.9 & 5.2 && 180 & 0.943 & 192.2 & 15.6 \\
23.75  && 23 &  0.806 & 25.6 & 5.8 && 177 & 0.938 & 173.4 & 15.5 \\
24.25  && 25 &  0.369 & 63.5 &14.0 && 263 & 0.842 & 299.7 & 20.5 \\
24.75  &&  5 &  0.069  &\nodata&\nodata&& 204& 0.466 & 419.9 & 32.2 \\
25.25  &&  0 &  0.004  &\nodata&\nodata&&  51 & 0.116 &\nodata&\nodata\\
\enddata
\tablecomments{The $V$ columns reports the $V_{\rm in}$ ($V_{\rm true}$) magnitude of 
the center of the 0.5 mag bins. $N_{\rm obs}$ are the observed number of stars 
in the bin. $C_f$ is the completeness factor in the center of the bin.
$N_c$ is the corrected number of stars in the bin (i.e. this column contains the
final corrected LF) and $\sigma_{Nc}$ is the  error on $N_c$. 
The final corrected LFs are reported only down to the first bin with 
$C_f<50 \%$.}
\end{deluxetable}



\begin{thebibliography}{300}
%
\bibitem[Andreuzzi et al.(2001)]{glo} Andreuzzi, G., De Marchi, G., Ferraro,
         F.R., Paresce, F., Pulone, L., Buonanno, R. 2001, A\&A, 372, 851
\bibitem[Bedin et al.(2001)]{bedin} Bedin, L.R., Anderson, J., King, I.R.,
         \& Piotto, G. 2001, \apj, 560, L75
\bibitem[Bellazzini et al.(2001a)]{mike} Bellazzini, M., Ferraro, F.R.,
         Fusi Pecci, F., Galleti, S., Catelan, M., Landsman, W.B. 2001a, \aj,
         122, 2569
\bibitem[Bellazzini et al.(2001b)]{miki} Bellazzini, M., Fusi Pecci, F.,
         Messineo, M., Monaco, L., Rood, R.T. 2001b, \aj, in press (Pap~I)
\bibitem[Catelan et al.(2001)]{cat} Catelan, M., Bellazzini, M., Landsman,
       W.B., Ferraro, F.R., Fusi Pecci, F., Galleti, S. 2001, \aj, 122, 3171
\bibitem[Djorgovski(1993)]{dj93} Djorgovski, S.G. 1993, in Structure and
         Dynamics of Globular Clusters,  S.G. Djorgovski and G. Meylan eds.,
         ASP, S. Francisco, ASP Conf. Ser., vol. 50, p. 373
\bibitem[Dinescu, Girard \& van Altena(1999)]{dana} Dinescu, D.I., Girard,
        T.M., van Altena, W.F. 1999, \aj, 117, 1792
\bibitem[Drukier et al.(1988)]{druk} Drukier, G.A., Fahlman, G.G., Richer,
        H.B., \& VandenBerg, D.A., 1988, \aj, 95,1415 	
\bibitem[Harris(1996)]{harris} Harris, W.E. 1996, \aj, 112, 1487
\bibitem[King et al.(1998)]{king} King, I.R., Anderson, J., Cool, A., Piotto,
         G. 1998, \apj, 492, L37
\bibitem[Lejeune et al. (2001)]{lej} Lejeune, T., \& Fernandes (eds.), J., 2001, 
         Observed HR diagrams and stellar evolution: the interplay between 
	 observational constraints and theory, 
	 ASP, S. Francisco, ASP Conf. Ser., in press
\bibitem[Marconi et al.(1998)]{marc} Marconi, G., Buonanno, R., Carretta, E.,
        Ferraro, F.R., Montegriffo, P., Fusi Pecci, F., De Marchi, G.,
        Paresce, F., Laget, M. 1998, \mnras, 293, 479
\bibitem[Gnedin \& Ostriker(1997)]{oleg} Gnedin, O., Ostriker, J.P. 1997,
        \apj, 474, 223
\bibitem[Meylan \& Heggie(1997)]{mh97} Meylan, G., \& Heggie, D. C. 1997,
       A\&A Rev., 8, 1
\bibitem[Mighell(1990)]{ken} Mighell, K.J. 1990, A\&AS, 82, 207
\bibitem[Ferraro et al.(1997)]{fer97} Ferraro, F.R., Carretta, E., Bragaglia,
         A., Renzini, A. Ortolani, S. 1997, \mnras, 286, 1012
\bibitem[Ferraro et al.(1999)]{fer99} Ferraro, F.R., Messineo, M.,
         Fusi Pecci, F., Straniero, O., Chieffi, A., \& Limongi, M. 1999a,
         \aj, 118, 1758
\bibitem[Pasquali, Brigas \& De Marchi(2000)]{anna} Pasquali, A., Brigas,
        M.S., De Marchi, G. 2000, A\&A, 362, 557 (PBD00)
\bibitem[Paresce \& De Marchi(2000)]{pdm} Paresce, F., De Marchi, G. 2000,
         \apj, 534, 870
\bibitem[Piotto \& Zoccali(1999)]{manu99} Piotto, G., \& Zoccali, M. 1999,
        A\&A, 345, 485 (PZ99)
\bibitem[Piotto \& Zoccali(2000)]{manu00} Piotto, G., \& Zoccali, M. 2000,
        A\&A, 358, 943 (PZ00)
\bibitem[Press et al. (1992)]{numrec} Press, W.H., Teukolsky, S.A., Vetterling,
         W.T., \& Flannery, B.P., 1992, Numerical recipes (2nd edition),
	 Cambridge, Cambridge University Press	
\bibitem[Pryor \& Meylan(1993)]{tad} Pryor, C., Meylan, G. 1993, in
       Structure and Dynamics of Globular Clusters,  S.G. Djorgovski and 
       G. Meylan eds., ASP, S. Francisco, ASP Conf. Ser., vol. 50, p. 357
\bibitem[Rosenberg et al.(1999)]{ros99} Rosenberg, A., Saviane, I., Piotto,
        G., \& Aparicio, A. 1999, \aj, 118, 2306
\bibitem[Rosenberg et al.(2000)]{ros00} Rosenberg, A., Piotto, G., Saviane,
        I., \& Aparicio, A. 2000, A\&AS, 144, 5
\bibitem[Rubenstein \& Bailyn(1997)]{ruba1} Rubenstein, E. P., \& Bailyn, C.
        D., 1997, \apj, 474, 701 (RB97)
\bibitem[Rubenstein \& Bailyn(1999)]{ruba2} Rubenstein, E. P., \& Bailyn, C.
        D., 1999, \apj, 513, L33 (RB99)
\bibitem[Shetrone \& Keane(2000)]{sk00} Shetrone, M. D., \& Keane, M. J.
       2000, \aj, 119, 840
\bibitem[Stetson \& Harris(1988)]{stha} Stetson, P.B., \& Harris, W.E., 1988,
        \aj, 96, 909      
\bibitem[Tosi et al.(2001)]{tosi} Tosi, M., Sabbi, E., Bellazzini, M., Aloisi,
         A., Greggio, L., Leitherer, C., Montegriffo, P., \aj, 122, 1271
\bibitem[Vesperini \& Heggie(1997)]{vh} Vesperini, E., Heggie, D.C. 1997,
         MNRAS, 289, 898

\end{thebibliography}
\end{document}